\documentclass[a4paper]{article}
\usepackage[utf8]{inputenc}
\usepackage[english]{babel}
\usepackage[T1]{fontenc}
\usepackage{natbib}																
\usepackage{amssymb,amsmath,amsthm}        
\usepackage{color}
\usepackage{xcolor}
\usepackage{graphicx}	
\usepackage{rotating}
\usepackage[a4paper,top=2cm,bottom=2cm,left=2cm,right=2cm,marginparwidth=1.5cm]{geometry}
\usepackage{amsmath}
\usepackage{graphicx}
\usepackage[colorlinks=true, allcolors=blue]{hyperref}

\title{A Probabilistic Formulation of the Diffusion Coefficient in Porous Media as Function of Porosity}

\author{Alraune Zech, Matthijs de Winter}
\date{\today}

\begin{document}

\maketitle

\section*{Abstract}

We investigate the upscaling of diffusive transport parameters as function of pore scale material structure using a stochastic framework. We focus on sub-REV (representative elementary volume)  scale where the complexity of pore space geometry leads to a significant scatter of transport observations. We study a large data set of sub-REV measurements on porosity and transport ability being a dimensionless parameter representing the ratio of diffusive flow through the porous volume and through an empty volume.
We characterize transport ability as probability distribution functions (PDFs) of porosity capturing the effect of pore structure differences among samples. We then investigate domain size effects and predict the REV scale. 
While scatter in porosity observation decrease linearly with increasing sample size, the observed scatter in transport ability converges towards a constant value larger zero. Our results confirm that differences in pore structure topology impact transport parameters at all scales. Consequently, the use of PDFs to describe the relationship of effective transport coefficients to porosity is advantageous to deterministic semi-empirical functions. We discuss the consequences and advocate the use of PDFs for effective parameters in both continuum equations and data interpretation of experimental or computational work. We believe that the presented statistics-based upscaling technique of sub-REV microscopy data provides a new tool in understanding, describing and predicting macroscopic transport behavior of micro-porous media.


\section{Introduction}

Characterizing flow and solute transport in porous media are essential for numerous applications in earth science, engineering and industry, including CO$_2$-storage, drinking water protection, safe disposal of nuclear waste, and enhanced oil recovery. 
Diffusion limited transport mechanisms, such as in subsurface gas transport in soils \citep{Jayarathne20}, preparative chromatography \citep{SchultzeJ20} and heterogeneous catalysis \citep{Ertl08}, depend on available pore space i.e.~porosity.

While the physics of transport are well understood, the complex geometry of natural materials poses a challenge for determining bulk transport parameters given the spatial heterogeneity of fluxes. 
Pore space is heterogeneous across large scales leading to observations being below the representative elementary volume (REV) level of the porous media process \citep{Bear72}. 
Even for Fickian diffusion, the impact of the complexity of the pore space on transport behavior is not fully captured in a generic mathematical framework. Or as \cite{Bruckler89} put it: \textit{"It appears that there is not a simple and unique relationship between the gas diffusion coefficient and air-filled porosity. [...] Predicting the gas diffusion coefficient on a large range of soil samples without complete measurements will probably be quite complicated."}

Relating transport properties at Darcy (or even Field) scale to structural parameters at pore scale has a long tradition starting with \cite{Hazen93} or \cite{Kozeny53} and many others. The complications of these relations are reflected by the great variety in (semi-empirical) functions throughout the literature \citep{VanBrakel74, Shen07}. Early attempts primarily involved the bulk porosity as the main determining parameter, later complimented by additional geometrical parameters, such as tortuosity, constrictivity, a matrix or formation factor, grain shape factors, etc. \citep{SchultzeJ20}. 
It remains a challenge to determine their individual contributions to transport properties at the Darcy scale. Hence, they are often used as fitting parameters in deterministic functional relationships ignoring the natural scatter in the data due to geometrical heterogeneity of porous material \citep{Ghanbarian13}.

The development of imaging techniques such as micro-computed X-ray tomography \citep{Cnudde13} and electron microscopy \citep{Grunwaldt13} boosted the development of digital rock physics (DRP) to directly calculate petrophysical properties \citep{Blunt13}. However, the complex pore geometry requires a sufficient sample number or large sample size to guarantee that results are representative in terms of volume and details captured.
While the heterogeneity of soils has triggered the development of stochastic methods in describing flow and transport at field scale decades ago \citep{Dagan89, Koltermann96, SanchezV06}, the development of a detailed statistical analysis of pore scale data sets on porous media transport in complex geometries has just recently started \citep{Mehmani20,Karimpouli16, deWinter16, Jiang13}. Capturing effects of material heterogeneity on transport parameters at sub-REV scale remains a challenge.

We study the upscaling of diffusive transport parameters using a stochastic framework to include the complexity of pore space geometry. We rely on a large data set from \citet{deWinter16} containing porosity values and diffusive transport observations of sub-REV domains. 
We show that the diffusion coefficient in porous media is not a deterministic function of pore structure and porosity but is better represented by a probability distribution. The stochastic representation of pore structure differences allows investigating domain size effects of observed scatter as well as making predictions on the scale of the representative elementary volume (REV). We further discuss implication on experimental work, continuum equations and simulations of large scale processes in the context of the REV concept. 

\section{Diffusion in Porous Media}
Diffusion is the net effect of Brownian motion across a region with a concentration gradient. Diffusion in porous media is classically described by a modified form of Fick's Law, relating diffusive flux to the concentration gradient:
\begin{equation}\label{eq:Fick01}
    J = D_\text{PM} \frac{d C(x)}{dx} 
\end{equation}
The proportionality constant $D_\text{PM}$ is interpreted as an effective diffusion coefficient being a lumped parameter containing: 1) all the dynamics and interactions of the fluid molecules (e.g.~molecular mass, molecular size, electrostatic interactions and thermodynamic properties); 2) the impact of the pore structure; and 3) all interactions between the molecules and the porous material. 
Consequently, the effective diffusion coefficient $D_\text{PM}$ is a function of pressure, temperature and molecular diameter as well as a great number of geometrical properties such as porosity, tortuosity, constrictivity, etc. 

The use of effective diffusion coefficients for describing diffusive transport at Darcy scale is only valid when the pore scale process is at REV level, which is by definition the smallest domain size at which a particular property can be described by an effective parameter \citep{Bear72}. The REV size depends on the process being investigated, where structural and transport properties differ in REV scales \citep{Zhang00}: while porosity is a pure volume property, tortuosity and flow depend on the actual topology of the pore space having much higher REV scales. 

In the classical picture of diffusion in porous media the effective diffusion coefficient is simplified to $D_\text{PM} = \frac{\theta D_\text{mol}}{\tau}$, where $\theta$ is the effective porosity, $\tau$ is tortuosity and $D_\text{mol}$ is the molecular diffusion constant of the particular fluid phase occupying the pore space.
Extracting geometrical parameters, such as $\theta$ and $\tau$ from microscopy images of the pore space is not trivial. For example the total porosity as ratio of pore to total sample volume differs from the effective porosity, which does not contain pore space not contributing to flow. Tortuosity $\tau$ as measure for the actual length of flow path is almost impossible to determine experimentally. 

We will make use of a dimensionless parameter representing the effect of the pore structure on diffusive flux: the transport ability $ta$. We define it as the ratio of diffusive flow through the porous volume and the flow through an empty volume (sec.~\ref{sec:defta}). In the classical diffusion picture $ta$ simplifies to ${\theta}/{\tau}$ and $D_\text{PM} = ta \cdot D_\text{mol}$.

\section{Observation Data and Statistical Description}

We analyse the steady-state diffusive transport in porous media following observations in Fluid Catalytic Cracking (FCC) particles. \cite{deWinter16} presented an extended data set of porosity observations for two types of FCC particles, which have an average diameter of $100\,\mu$m micrometer and pore sizes ranging from $1$\,nm to roughly $2\,\mu$m. Cross sections of entire FCC particles were examined with a focused ion beam-scanning electron microscope (FIB-SEM) to determine the spatial distribution of the porosity. The 3D pore space geometry was determined by FIB-SEM tomography.

\subsection{Porosity}
\label{sec:pordata}
Porosity, the ratio of pore volume to total volume, has been determined for sub-samples at $2\,\mu$m, $8\,\mu$m, and $32\,\mu$m resolution \citep{deWinter16}. Although observations refer to areal investigation, we consider them as representative for the volume property, assuming isotropy in the third dimension. 

\begin{table}[ht]
    \centering
\begin{tabular}{|l| c c c | c c c |}
     \hline
                                            & \multicolumn{3}{l|}{FCC1} & \multicolumn{3}{l|}{FCC2} \\
     domain length/resolution $r$ [$\mu$m]  &   2x2 & 8x8  & 32x32 &   2x2 & 8x8  & 32x32     \\ 
     \hline
     mean porosity $\mu_r$              &   0.241 & 2.39 & 2.39 & 0.296 & 2.9 & 2.8     \\  
     standard deviation $\sigma_r$   &   0.074 & 0.048 & 0.001 & 0.145 & 0.09 & 0.044   \\
     \hline
\end{tabular}
    \caption{Statistical quantities of porosity observations for FCC1 and FCC2 catalytic particles as function of sample resolution $r = 2, 8, 32\mu$m.}
    \label{tab:porstats}
\end{table}

\begin{figure}[ht]
\centering
    \includegraphics[width=0.5\textwidth]{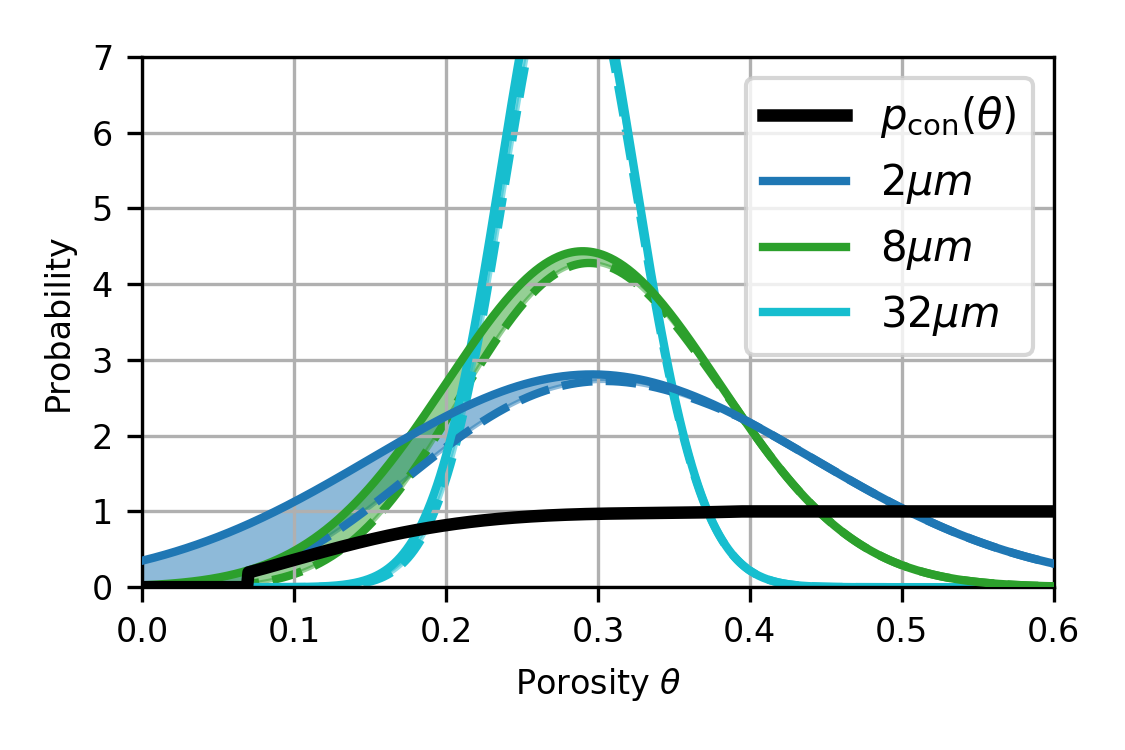}
\caption{Distribution function of porosity depending on resolution $r=2, 8, 32 \mu$m. Solid lines show truncated Gaussian PDF of total porosity. Black line shows the connectivity probability $p_\mathrm{con} (\theta) \in [0,1]$. The dashed lines indicate the distribution of connected porosity and the shaded area between the curves indicate the amount of non-connected porosity's.}
    \label{fig:porosity}
\end{figure}

Observed porosities of individual material samples show a clear normal distribution. Statistical analysis of samples lead to scale dependent mean and standard deviations as listed in Table~\ref{tab:porstats}. The probability of observing a certain porosity $\theta$ in a samples of resolution $r$ can be described with a truncated normal distribution: $ P^{(r)}(\theta)\propto N(\mu_r,\sigma_r)$:
\begin{equation} \label{eq:por_normal}
    P_\theta^{(r)}(\theta = x) = \frac{1}{\sigma_r\sqrt{2\pi}} \exp{\left(-\frac{(x-\mu_r)^2}{2 \sigma^2_r}\right)}
\end{equation}
to the parameters listed in Table~\ref{tab:porstats}. Note the truncation of values to the range of $\theta \in [0,1]$. Figure~\ref{fig:porosity} shows the truncated normal distributions of porosity's at all resolution levels for the FCC2 material.

Porosity statistics show a scale dependency according to volume averaging impacted by spatial correlation. As expected, mean porosity $\mu$ is quasi constant, whereas the standard deviation~$\sigma$ decreases with increasing resolution: $\sigma_{32} \approx 0.5 \sigma_8 \approx 0.25 \sigma_2$. $\sigma$ scales with the domain length increase $n = r_{i+1}/r_i$ with $1/\sqrt{n}$ which is the root of the rate predicted by the law of large numbers for independent samples. Note that this analysis relates to 2D, where an increase of domain length by $n=4$ units correspond to a network consisting of $n^2=16$ sub-elements. Sampling in 3D increases the total number of network sub-elements to $n^3=64$.

The slow decrease of the standard deviation shows that samples are not independent due to spatial correlation given the non-uniform and non-random material structure innate to porous material. The non-zero value of the standard deviation at $\sigma_{(32)}$ shows another effect: domain size has not yet reached the (porosity) REV level. The slower convergence of the standard deviation to zero increases the (porosity) REV scale since deviations of porosity observations can still be present at relatively large samples size. Both aspects, sub-REV scale and spatial correlation of samples need to be accounted for in the upscaling process.

\subsection{Transport Ability}

\subsubsection{Definition}
\label{sec:defta}
When investigating the impact of the pore structure on diffusive transport, we define a dimensionless "transport ability" $ta$ as ratio of the flux in the porous medium domain, i.e.~a restricted flux $J_\text{PM}$, to the flux in free domain, i.e.~an unrestricted flux $J_0 = D_\text{mol} \frac{d C(x)}{dx} $: 
\begin{equation}\label{eq:ta01}
    ta = \frac{J_\text{PM}}{J_0} = \frac{J_\text{PM}}{D_\textup{mol} \cdot dC/dx}
\end{equation}
The definition~\eqref{eq:ta01} of $ta$ allow extracting the effect of porous medium structure on diffusive flux from fluid specific impact. Thus, $ta$ is a sole property of the porous media structure, and we consider it a function of porosity $\theta$ and the topology of the void space $ta = f(\theta, \text{topology})$. Note that $ta$ is scale dependent.

$ta$ ranges between zero and one. For unrestricted domains (i.e. $\theta \rightarrow 1$), we have $ta \rightarrow \theta/\tau = 1$. However, for strongly restricted domain, related to small porosities $\theta$ and highly tortuous pore space, $ta$ is small. $ta = 0$ covers the case of a disconnected medium, i.e. no flow path through the medium is available. 

$ta$ is not a biunique property of porosity since various structures with identical porosity show significantly different flux pattern, such as straight flow channels or disconnected void space. The effect of structure is usually lumped into an "effective porosity", while the portion of void space contributing to transport is particularly critical.

\subsubsection{Transport Ability Observations}

\begin{figure}[ht]
\centering
    \includegraphics[width=0.5\textwidth]{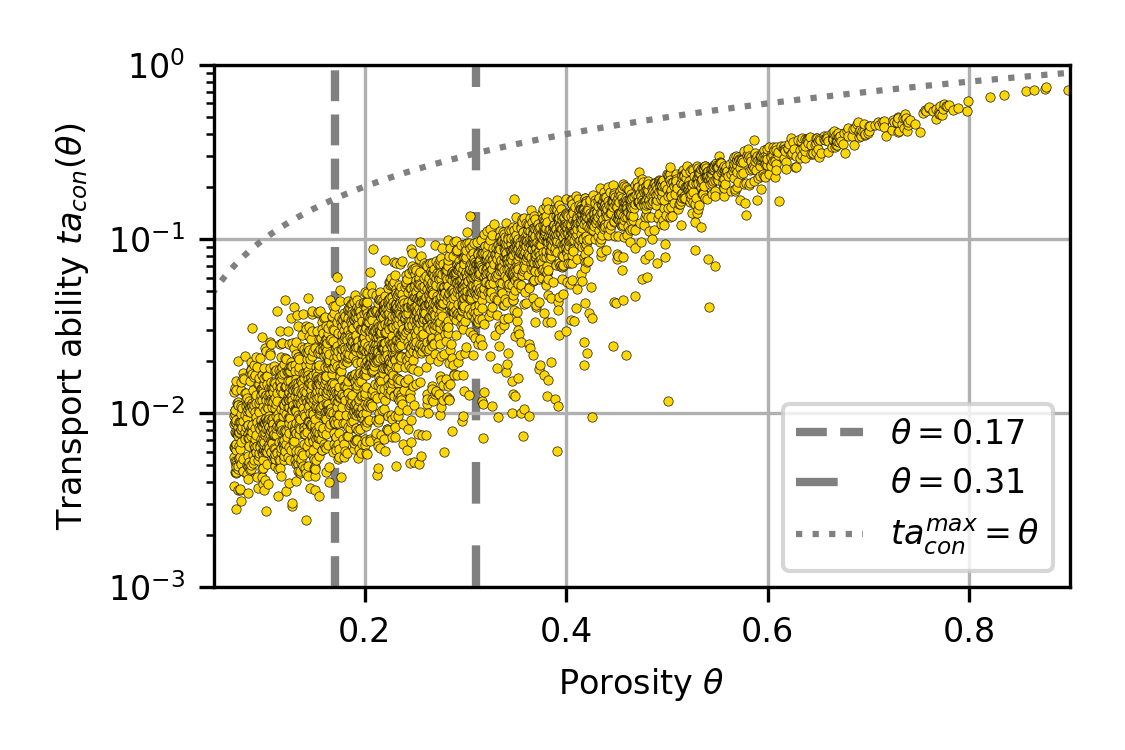}
\caption{Scatter plot of transport ability as function of porosity for connected volumes. Dashed line indicates maximal $ta$ corresponding to plug flow. Vertical gray lines indicate positions of individual $ta(\theta)$ analysis in Figure~\ref{fig:histogram}.}
    \label{fig:scatter_data}
\end{figure}

We use a data set of transport ability observations form $5128$ cubic volumes of $r=2 \mu m$ length. Results are based on diffusive transport simulations for virtual volumes with a computer-generated pore space mapping the structure of the FFC particles. For details we refer to \cite{deWinter16}. To derive the probabilistic relation of transport ability $ta$ to porosity $\theta$, each volume $i$ is analyzed with respect to: (i) porosity $\theta_i$, (ii) connectivity, i.e. if a flow path through the sample exist to allow diffusive flux, (iii) transport ability $ta_i$ for the connected samples. We first focus on connected volumes for which the distribution of observed transport ability $ta_i$ versus porosity $\theta_i$ is displayed in Figure~\ref{fig:scatter_data}.

The data scatter in Figure~\ref{fig:scatter_data} shows that the domain is at sub-REV scale for both, $ta$ and $\theta$. Volumes of the same $\theta$ show large variations in $ta$, ruling out a deterministic one-to-one relation between both, certainly at sub-REV level. Consequently, we model $ta$ as random function of $\theta$ with a probability distribution function (PDF) $P_{ta}(\theta)$ representing the scatter of $ta$ for identical porosities. 

\subsubsection{Descriptive PDF for Transport Ability}
\label{sec:ta_con}

\begin{figure}[ht]
\centering
    \includegraphics[width=0.9\textwidth]{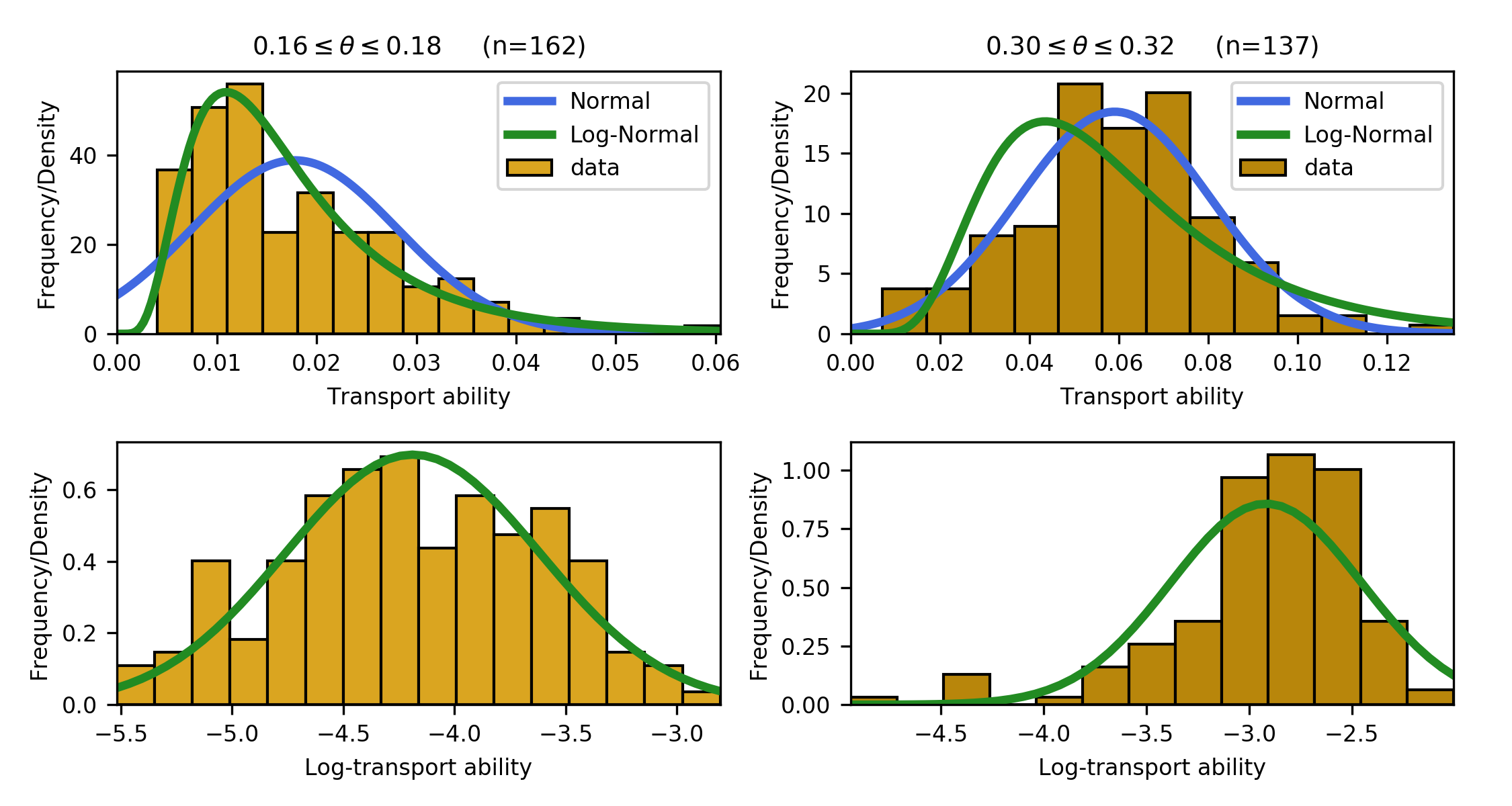}
\caption{Statistical analysis of transport ability for individual porosity values: Normalized frequency distributions of data in normal (top) and log-scale (bottom) compared to normal and log-normal distribution for $\theta = 0.17$ and $\theta = 0.31$. $n$ is the number of samples.}
    \label{fig:histogram}
\end{figure}

We examine the frequency of $ta$ values for small ranges of porosity values (e.g. $0.15 < \theta < 0.17$) as displayed in Figure~\ref{fig:histogram}. Details of the statistical analysis, including normality and log-normality tests are discussed on the \textit{Supporting Information}. 

For each porosity value, the $ta$ values follow a log-normal distribution which changes in shape depending on $\theta$. $ta$ distributions for small $\theta$-values show a strong log-normal shape while distributions become less skewed with increasing porosity (Figure~\ref{fig:histogram}). We attribute this tendency to the impact of pore space topology. The scarce pore space at small porosity value is either scattered with several bottlenecks or forms one or few flow channels. In the first domain type, flow is hampered and $ta$ will be low, while in the latter we observe quasi plug flow with a relatively high $ta$ values. The effect reduces with increasing void volume.

\begin{figure}[ht]
\centering
    \includegraphics[width=0.9\textwidth]{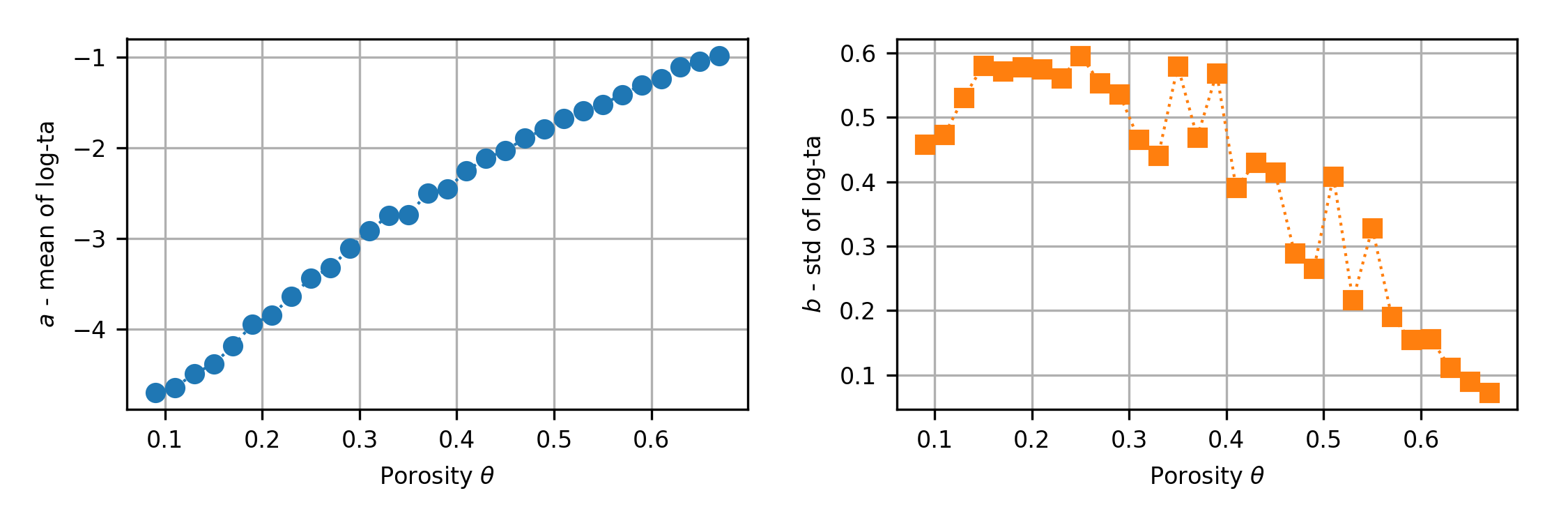}
\caption{Expectation values (mean) $a(\theta)$ and standard deviation $b(\theta)$ of log-transformed transport ability as function of porosity.}
    \label{fig:ta_mean_std}
\end{figure}

We calculated expectation values $a_{\log ta}$ and standard deviations $b_{\log ta}$ for each log-transformed frequency distribution. The log-transformation allows comparing statistics given the logarithmic nature of $ta$ values. The log-normal distributions also represents the frequency for higher porosities, as it converges to the normal-distribution for decreasing standard deviations. Results are displayed in Figure~\ref{fig:ta_mean_std}. 

The log-$ta$ mean $a(\theta)$ shows a linear relationship for porosities smaller then $0.6$ and then a flattening towards $0$ since $ta(\theta=1) = 1$. $\log-ta$ standard deviation is high for small porosities and decreases with increasing $\theta$. For porosities beyond $0.6$, the scatter in $ta$ is negligible. 

The parameters $a_{log-ta}$ and $b_{log-ta}$ can be used to encapsulate the dependency of $ta$ on porosity $\theta$ through functional description of the probability distribution  $P_\mathrm{ta}^\mathrm{con} = LN(a(\theta),b(\theta))$:
\begin{equation} \label{eq:ta_lognormal}
    P_\mathrm{ta}^\mathrm{con}(ta = y,\theta) = \frac{1}{yb(\theta)\sqrt{2\pi}} \exp{\left(-\frac{\left(\ln (y) -a(\theta)\right)^2}{2 b^2(\theta)}\right)}
\end{equation}
$P_\mathrm{ta}^\mathrm{con}(ta = y,\theta)$ describes the probability of $ta=y$ given a specific value of porosity $\theta$ assuming that the domain is connected, i.e. having a flow path through the sample. However, disconnected elements exist and have a transport ability of zero. 

\subsubsection{Connectivity Statistics} 
Connectivity is a sub-property of transport ability and similarly depends on porosity: decreasing pore volumes lower the probability of having a void space path through the domain. We model connectivity as probabilistic Bernoulli variable $X$ being either connected ($ta\neq0$) or not-connected/dead-end ($ta=0$) with the connectivity probability $p_{con}(\theta) \in [0,1]$ as function of porosity $\theta$: $P(ta\neq1)=p_\mathrm{con}(\theta)$ and $P(ta=0)=1-p_\mathrm{con}(\theta)$. 

The frequency analysis of the $5128$ virtual volumes with regard to connectivity resulted in a function $p_{con}(\theta)$ as displayed in Figure~\ref{fig:porosity}.
Samples of porosity smaller then $0.07$ are almost never connected $p_{con}(\theta<0.07)=0$, whereas samples of $39\%$ porosity or larger are always connected, $p_{con}(\theta>0.39)=1$. For porosities in between, \cite{deWinter16} determined a the functional description:
    \begin{equation}\label{eq:percprob}
            p_{con}(\theta) = 
            \begin{cases}
                    0  & \text{for } \quad\quad \theta \leq 0.07\\
                    \left(c_3\theta^3+c_2\theta^2+c_1\theta+c_0\right)^2 & \text{for }  0.07 < \theta < 0.39 \\
                    1  & \text{for } 0.39 \leq \theta
            \end{cases}
    \end{equation}
with $c_3=27.362$, $c_2=-27.661$, $c_1=9.4256$, and $c_0=-0.0927$. 

Relating the porosity distribution and the probability of connectivity in an ensemble allows determining the total amount of connected and disconnected samples. The dashed lines in Figure~\ref{fig:porosity} indicate the distribution of connected porosities. The shaded areas between the curves are the amount of dead-end samples with zero transport ability: $1- \int_0^1 P_r(\theta) \cdot p_\mathrm{con} (\theta) d\theta$. The number decreases with increasing samples resolution $r$ given the reduction of standard deviation $\sigma_r$ and thus less samples of small porosity. For the FCC2 material, the total amount of disconnected samples is predicted with $14.8\%$ for $r=2\mu$m; $9.7\%$ for $r=8\mu$m; and $6\%$ for $r=32\mu$m. Note that the number of disconnected samples is still significant for the sample size of $r=32\mu$m although the standard deviation is small and most samples range around the mean porosity of about $0.3$.

Transport ability $ta$ as stochastic function of porosity follows from the statistical descriptions of connectivity (Eq.~\ref{eq:percprob}) and connected transport ability (Eq.~\ref{eq:ta_lognormal}) with
\begin{equation}\label{eq:prob_ta}
P_\mathrm{ta}(ta = y,\theta) = 
\begin{cases}
    0  & \text{ with } 1-p_\mathrm{con}(\theta) \\
    y>0 & \text{ with } p_\mathrm{con}(\theta)\cdot P_\mathrm{ta}^\textup{con}(y,\theta)
\end{cases}
\end{equation}

Moments, such as expectation value and variance of $ta$ as function of $\theta$ can be derived based on the characteristics of the log-normal distribution. Theoretical expressions are listed in the \textit{Supporting Information}. Regard that moments refer to ergodic conditions, i.e. when the domain size is above REV level for all involved processes. 

\section{Upscaling}

We perform upscaling of observations addressing the questions of interest: To what extend is the scatter of transport ability $ta$ and porosity $\theta$ related to domain size? And at what domain size can we consider the REV to be reached? The evolution of porosity statistics for observations at some increasing domain size are outlined in section~\ref{sec:pordata}. Since observations of transport ability are not available at domain length beyond $2\,\mu$m, we simulated the scaling behaviour of $ta$ with increasing domain size. Based on these results we derive statistical relations between $ta$ and $\theta$ as function of the domain size~$r$.

\begin{figure}[htb]
\centering
    \includegraphics[width=0.55\textwidth]{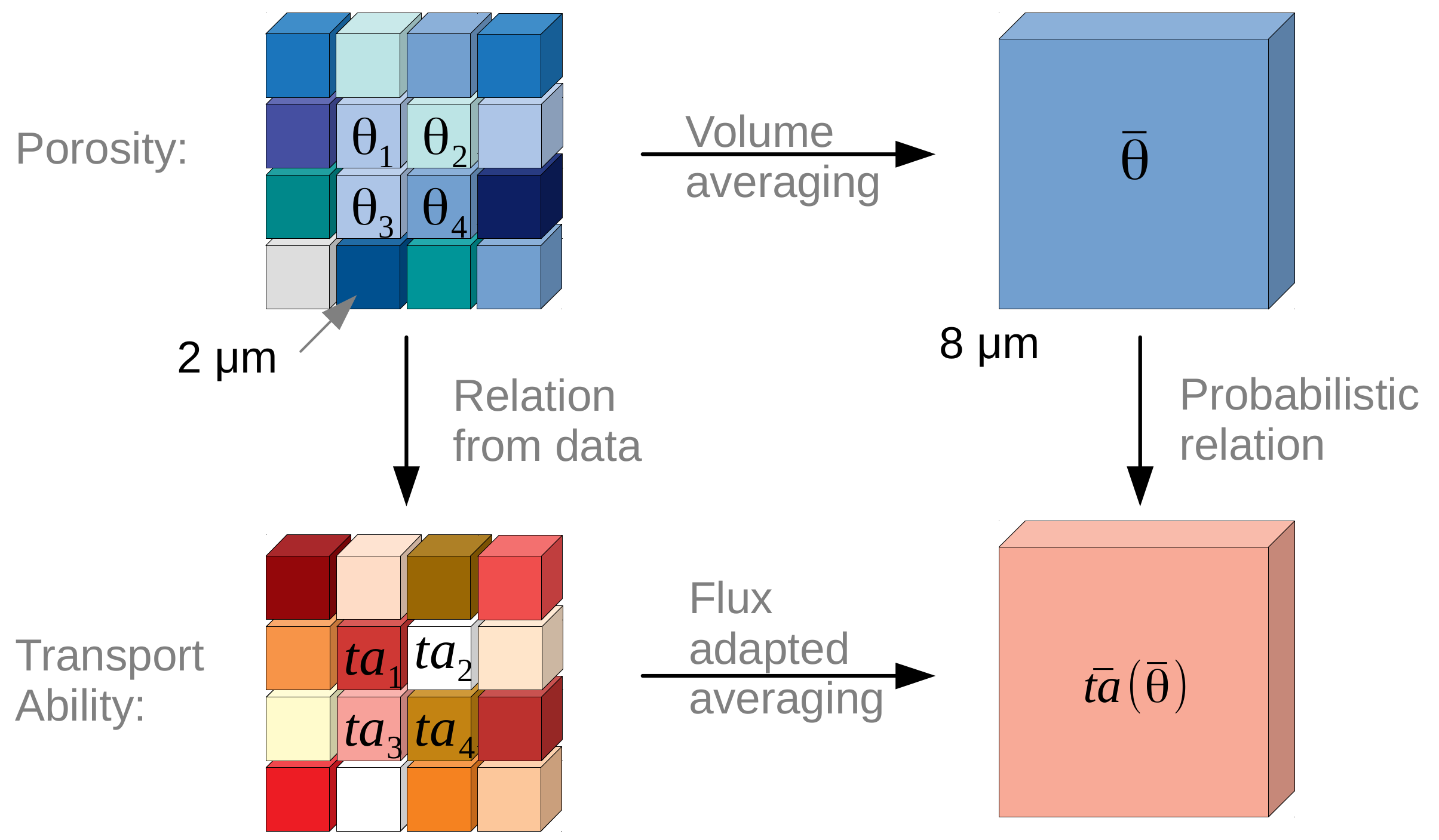}
    \caption{Upscaling Scheme: from known porosity distribution at small scale (upper left corner) to desired description of transport ability $\bar {ta}(\bar \theta)$ at larger scale (lower right corner). We determine $\bar {ta}(\bar \theta)$ numerically from upscaling small scale transport ability through flux adapted averaging (lower left corner) and derive a probabilistic relationship from porosities at larger scale $\bar \theta$ (upper right corner).}
    \label{fig:upscaling}
\end{figure}

We follow an upscaling procedure exemplified in Figure~\ref{fig:upscaling}. We determine transport ability at increasing scales within sub-REV level making use of numerical diffusive flux simulations in a Monte Carlo setting. We study the upscaling behaviour in 2D and 3D, since porosity measurements are often taken in a 2D setting, while flow and thus transport ability generally requires observation in 3D volumes. 

\subsection{Numerical Upscaling}

We generate discrete networks each consisting of $N = n^d$ sub-samples, with~$d$ being the dimension. We decided for $n=16$ elements per direction based on network theory and preliminary tests showing that boundary effects are negligible for $n>10$ (\textit{Supporting Information}). For instance a domain at $r=32\,\mu$m resolution consists of $16$ sub-samples of length $2\,\mu$m within each dimension. An exception are networks at resolution $r=8\,\mu$m which are upscaled from $n=4$ elements of length $2\,\mu$m.   

We generate ensembles of $10,000$ networks with domain resolutions between $r=8\,\mu$m and $r=2048\,\mu$m. All networks of an ensemble share the same statistics and resolution. Tests on ensemble convergence showed that the ensemble size of $10,000$ is sufficient to achieve reproducible results. 

The workflow for generating one network within an ensemble comprises:
\begin{itemize}
    \item The network is initialized with $N = n^d$ elements. $\sqrt{N}$ random porosity values $\theta_i$ are drawn from a truncated normal distribution $P_\theta(\theta)$ (Eq.~\ref{eq:por_normal}) with mean $\mu=0.3$ and $\sigma_r$ according to the resolution adapted statistics based on the values determined for the FCC2-1 material (Table~\ref{tab:porstats}). Each porosity value is assigned to $\sqrt{N}$ neighboring nodes in the network. The porosity generation accounts for the spatial correlation of porosity and preserves the statistics observed in the data (section~\ref{sec:pordata}).
    \item Connectivity of each element is randomly specify as either yes($=1$) or no($=0$) based on its porosity $\theta_i$ and the associated probability of connectivity $p_\mathrm{con}(\theta_i)$ (Eq.~\ref{eq:percprob}), which we considered resolution independent. 
    \item A transport ability value $ta_i$ of each connected node is drawn from the probability distribution $P_\mathrm{ta}^\mathrm{con}(\theta_i)$ (Eq.~\ref{eq:ta_lognormal}) whose statistics $a$ and $b$ are based on the $ta$ value distribution of the domain size $r$ of the nodes. 
    Tests using a distribution function based in the histograms of $ta_r$ values instead of imposing a log-normal function showed identical results. 
\end{itemize} 

For each generated network we determine upscaled properties $\bar \theta$ and $\bar{ta}$. The average porosity $\bar \theta = \sum_{i} \theta_i$ is the arithmetic mean of the nodes' porosities. The network's effective transport ability $\bar{ta}$ follows from a flow simulation: (i) apply a pressure gradient to the network; (ii) solve the diffusive flux equation~\eqref{eq:Fick01} numerically by solving the matrix equation where the adjacency matrix is build according to the $ta$ values; (iii) calculate $\bar{ta}$ as ratio of the imposed concentration gradient and the calculated flux according to Eq.~\eqref{eq:ta01}. 

\begin{figure}[ht]
\centering
    \includegraphics[width=0.475\textwidth]{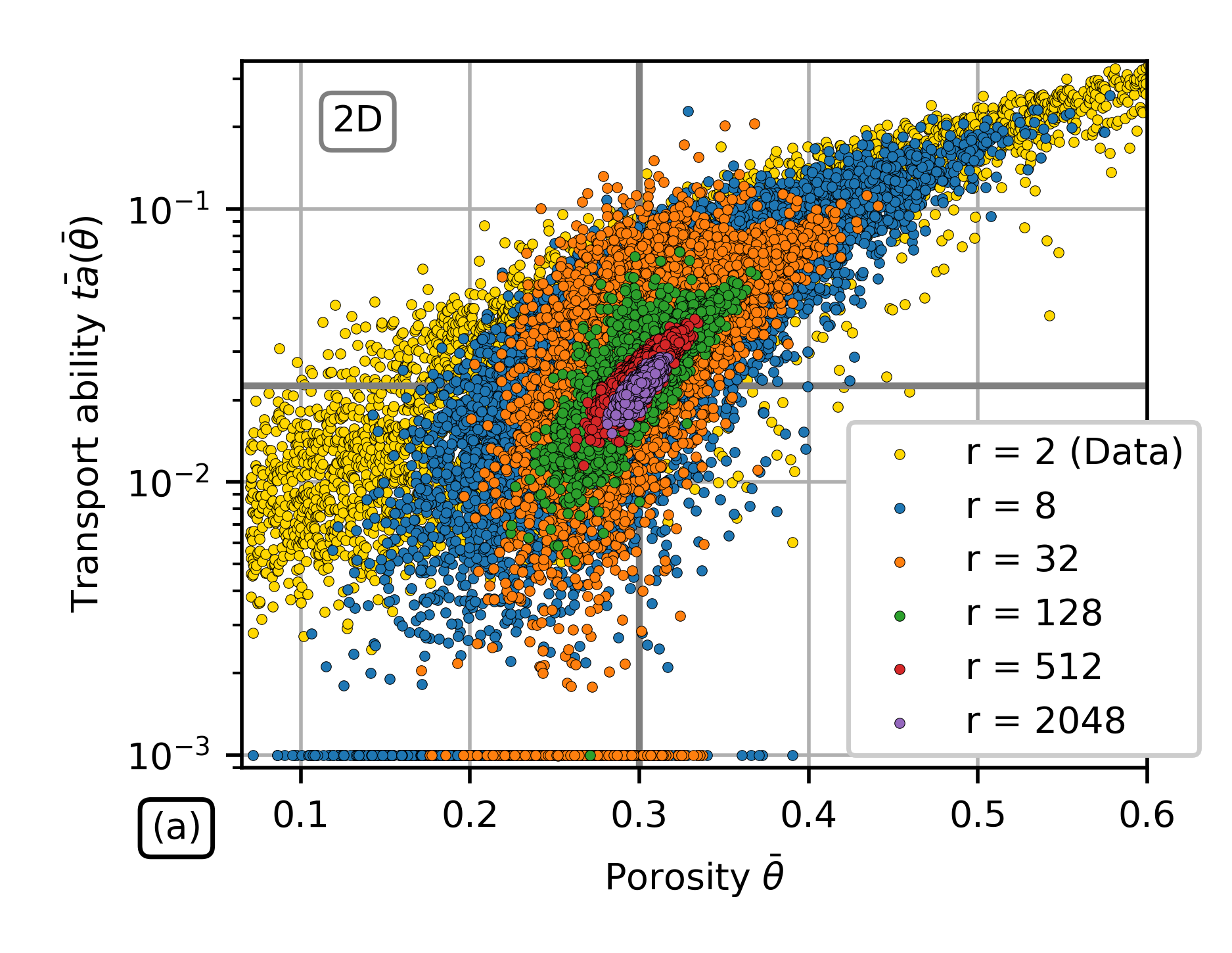}
    \includegraphics[width=0.95\textwidth]{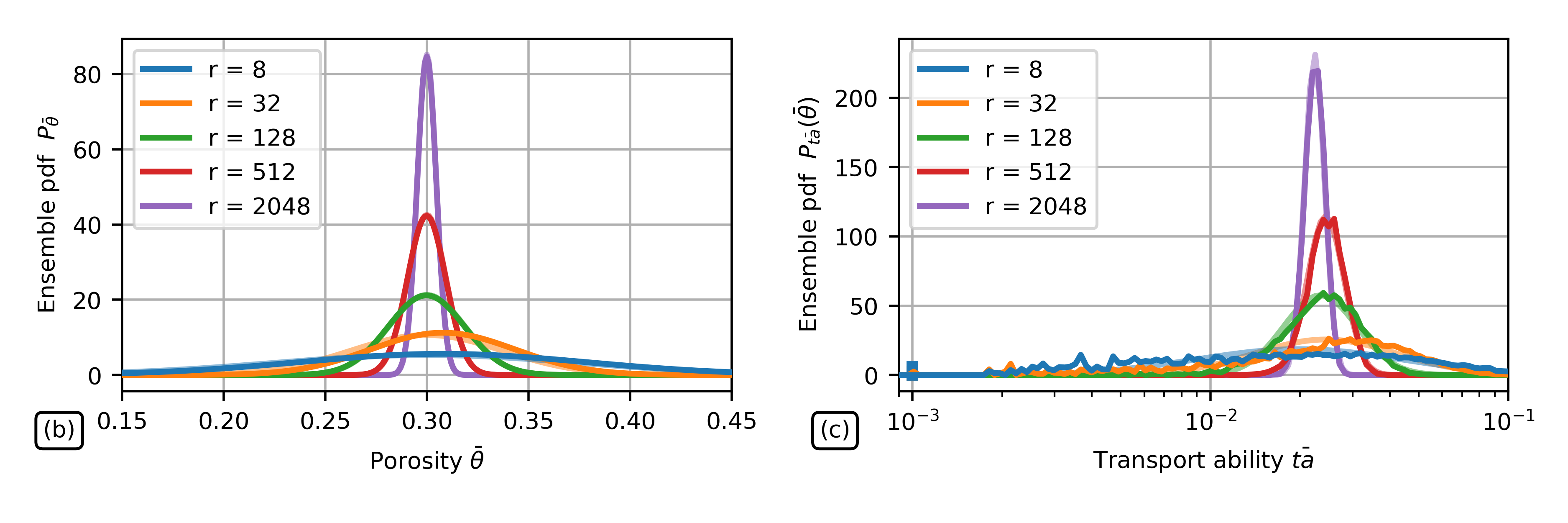}\caption{Results of numerical upscaling for simulated ensembles at different resolutions $r$ [$\mu$m] in 2D: (a) Scatter of ensemble transport ability $\bar{ta}$ vs porosity $\bar\theta$; (b) and (c) marginal distributions (i.e. normalized histograms) for porosity and transport ability. Lighter lines slightly visible in (b) and (c) show the associated theoretical distributions based ensemble parameters.}
    \label{fig:scatter_eff}
\end{figure}

\subsection{Numerical Upscaling Results}

Figure~\ref{fig:scatter_eff} shows the results on upscaled transport ability $\bar{ta}$ and porosity $\bar\theta$ for simulated ensembles at resolutions $r=8-2048\mu$m. Each dot in Figure~\ref{fig:scatter_eff}a represents one of the $10000$ networks in the ensemble. Displayed results refer to 2D, while similar results for the 3D setup are accessible in the \textit{Supporting Information}. Marginal distributions show the normalized frequencies from a histogram analysis of $\bar\theta$ and $\bar{ta}$ data within each ensemble. The marginal distributions confirm that porosity and log-transport ability follow normal distributions with scale dependent parameters.

Porosity distributions of each ensemble are normal distributed and perfectly in line with theoretical upscaling behaviour of the FFC particle data (section~\ref{tab:porstats}): constant mean $\mu$ and decreasing standard deviation $\sigma_r = 0.15/\sqrt[4]{N_r}$ where $N_r = (r/2)^d$ is the the total number of nodes of size $r_0=2\,\mu$m (the initial resolution) in each network of resolution $r$. 

The scaling behaviour of transport ability combines two effects: a decrease of disconnected networks and a decreasing mean in connected transport ability with increasing resolution. At a level of $r=512\,\mu$ basically all networks are connected and have transport ability $>0$. However, these values can be rather small and now contribute to the mean of connected transport ability, which thus decreases. Figure~\ref{fig:scatter_eff}c shows that a normal distribution matches well the $\log ta$ data (i.e.~log-normal for $ta$) particularly at increasing resolution level. Now, the distribution represents the entire $ta$ behaviour, since all elements are connected. Note that for small resolutions the impact of disconnected elements with $ta=0$ are not included in the mean $a$ of $\log ta$-values. The standard deviation $b$ of log-$ta$ decreases with scale.

\begin{figure}[ht]
\centering
    \includegraphics[width=0.49\textwidth]{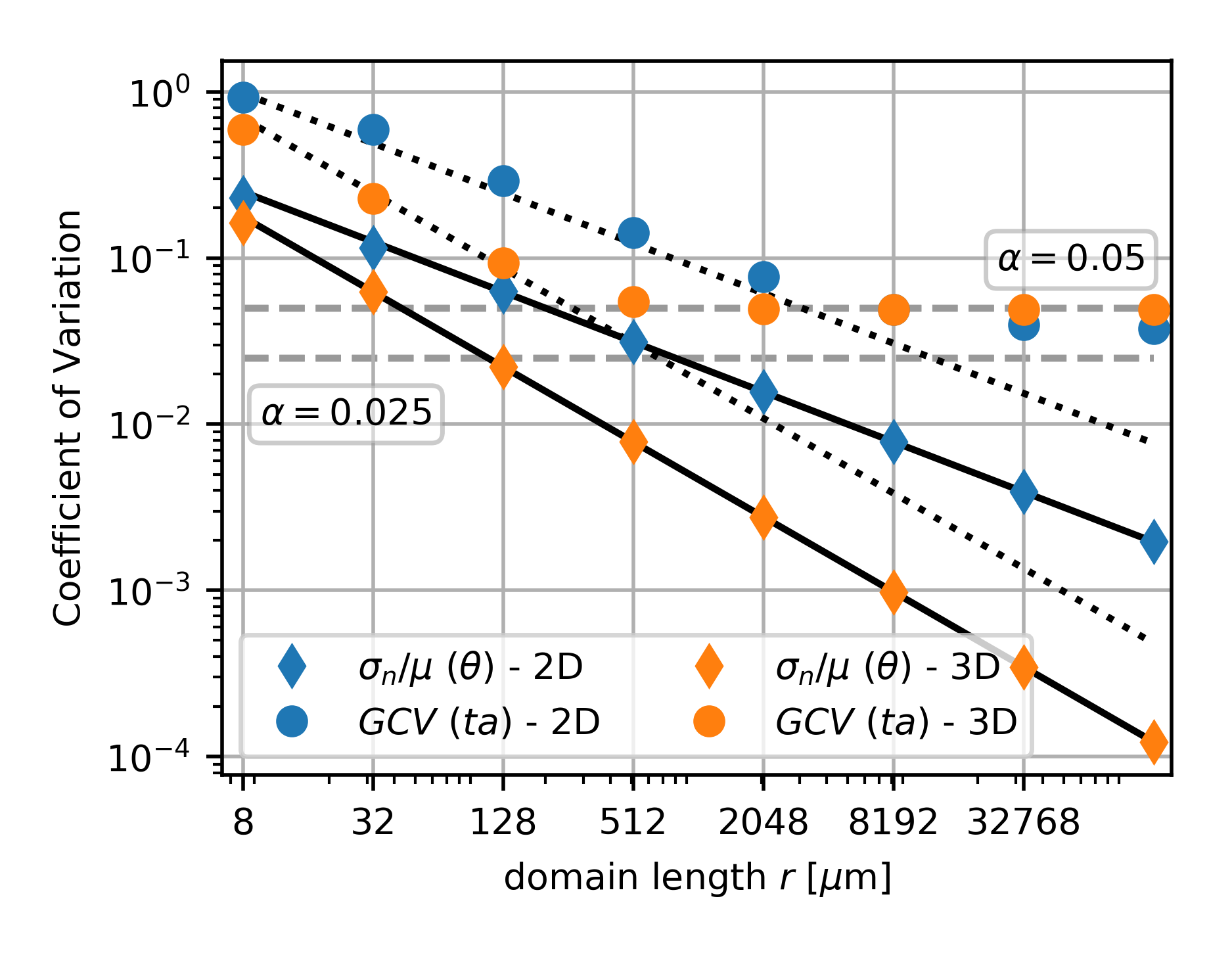}
    \includegraphics[width=0.49\textwidth]{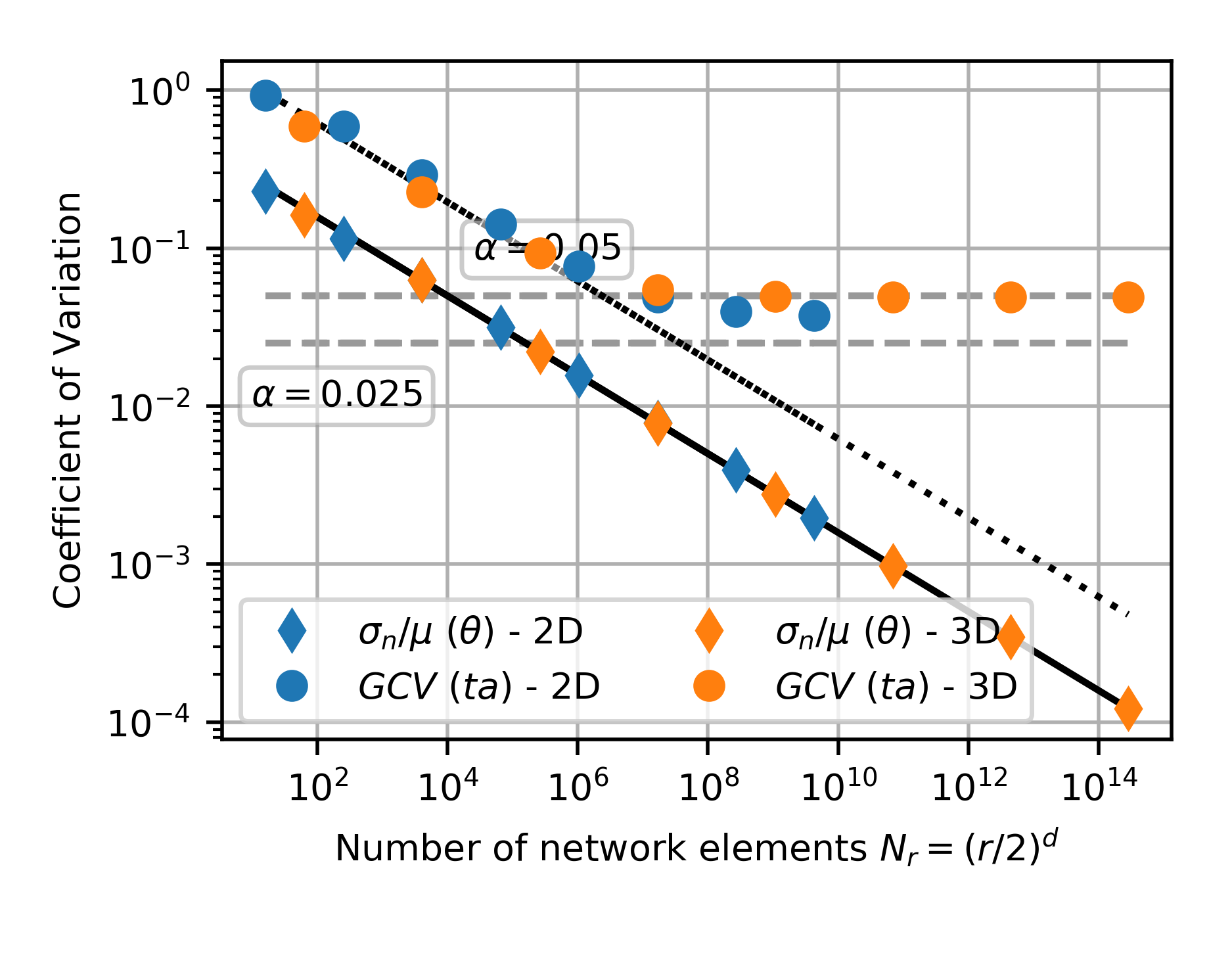}
    \caption{Evolution of porosity and transport ability statistics with increasing domain resolution $r$: Diamonds show coefficient of variation of porosity statistics of simulated ensembles $CV_{\bar\theta} = \sigma_r/\mu$, circles show geometric coefficient of variation $GCV_{\bar{ta}} = \sqrt{\exp{(b^2)} -1}$ for the transport ability. Lines show theoretical predictions by a scaling proportional to $1/\sqrt[4]{n^d}$. Left scaling according to domain resolution~$r$, right scaling to number of elements in network $N_r$.}
    \label{fig:ens_stats_evolution}
\end{figure}

We investigate the scale dependence of statistics of porosity $\bar \theta$ and transport ability $\bar{ta}$, particular that of the standard deviation characterizing the decreasing spread, making use of the coefficient of variation: $CV_{\bar\theta} = \sigma/\mu$ scales the standard deviation by mean. For $ta$ we make use of the geometric coefficient of variation $GCV_{\bar{ta}} = \sqrt{\exp{(b^2)} -1}$ being more appropriate for log-normal distribution. We associate a sufficiently low CV with reaching the REV level. For porosity e.g.~a $CV<0.025$ corresponds to a standard deviation of $2.5\%\mu$ and consequently more then $95\%$ of the porosity values in the PDF lie within a range of $\mu \mp 2\sigma = [1.05\mu,1.05\mu]$. 

Figure~\ref{fig:ens_stats_evolution} shows the scaling behaviour of the statistics of porosity and transport ability for both dimensional analyses, $2D$ and $3D$. The display of CV as function of domain resolution $r$ shows a steeper decrease in 3D due to the higher amount of sub-samples with the additional dimension. Differences disappear when displaying the scaling behavior depending on the total number of elements $N_r = (r/2)^d$, which is a function of the dimension.

Parameter scaling of porosity follows the theoretically predicted relation $CV_\theta = \sigma_r/\mu=\frac{0.15}{0.3\sqrt[4]{N_r}}$, which is $CV_\theta^{(2D)} = 1/2\sqrt{0.5r}$ and $CV_\theta^{(3D)} = 1/2(0.5r)^{3/4}$ in 2D and 3D, respectively. Both relations show a linear fit in log-log display in Figure~\ref{fig:ens_stats_evolution}. The $CV$ is below a level of $0.025$ for a resolution of $n=512\,\mu$. Thus, we consider the REV level for porosity reached at that resolution in a 2D sample analysis. 

Parameter scaling of transport ability, shows a similar trend with decreasing $GCV$ at a rate proportional to $1/\sqrt[4]{N_r}$. However, at large resolutions the trends flattens out and log-$ta$ standard deviation $b$ does not decrease any further. The fact that the spread of $\log-ta$, does not converge to zero can be explained with differences in pore structure at all scales, leading to a remaining scatter of $ta$ values for a specific porosity. The asymptotic value lies at about $\alpha = 0.05$. 

We relate the REV of $ta$ to the 3D upscaling behavior since flow is best determined in volumes. Assuming a REV level for $ta$ at a GCV level of $\alpha = 5\%$, this is reached at a resolution of $r=2048\mu$\,m. That corresponds to a domain volume of about $1e10$\,$\mu m^3$. Figure~\ref{fig:ens_stats_evolution} makes obvious that spatial resolutions for reaching REV levels differ significantly between porosity and transport ability.

\subsection{Theoretical Upscaling}

The statistical descriptions of porosity, connectivity and connected transport ability allows theoretical upscaling of $ta$ as stochastic function of porosity. We further make us of the findings of numerical upscaling to derive an expression for the stochastic description of transport ability as function of porosity~$\theta$ and domain resolution~$r$. 

\begin{figure}[ht]
\centering
    \includegraphics[width=0.95\textwidth]{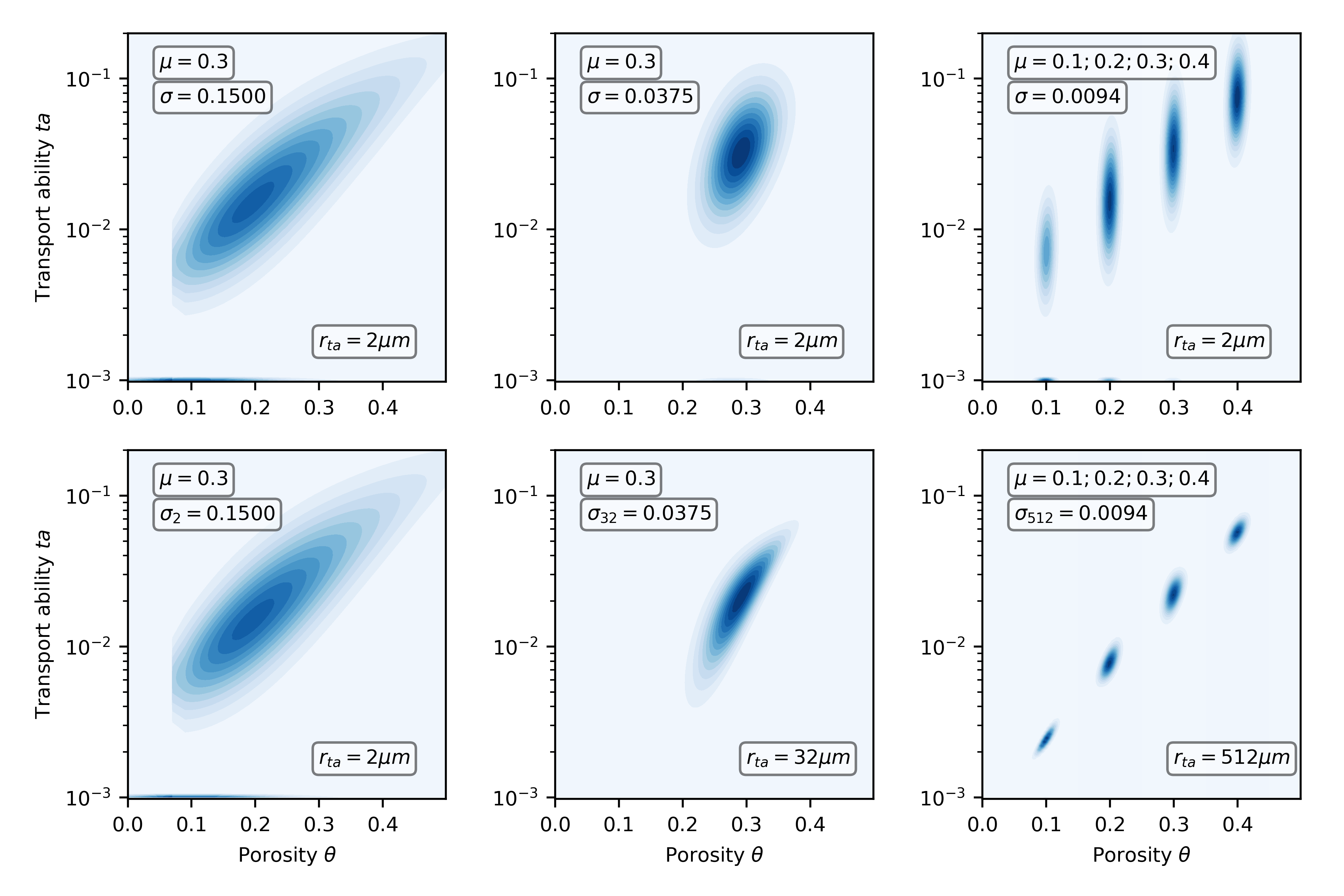}
\caption{Probability contours for transport ability $ta$ as function of porosity $\theta$ with various porosity mean $\mu$ and standard deviation $\sigma$ values (Eq.~\eqref{eq:por_normal}). 
Upper row has identical $ta$ statistics of $a_r$ and $b_r$ of log-normal distribution (Eq.~\ref{eq:ta_lognormal}) at $r=2$, while lower row $ta$ statistics refer to scales $r=2\mu$m, $r=32\mu$m and $r=512\mu$m which correspond to porosity standard deviations $\sigma_r$.}
    \label{fig:cloud_TA_pdf}
\end{figure}

Starting point is a material at resolution $r$ with a normal porosity distribution $P_\theta^{(r)}$ (Eq.~\ref{eq:por_normal}) of mean $\mu$ and standard deviation~$\sigma_r$. 
The distribution $P_\mathrm{ta}^{(r)}$ of connected transport ability at $r$ can be assumed as log-normal distributed to parameters $a_r$ and $b_r$ according to (Eq.~\ref{eq:ta_lognormal}). Non-connected elements distribute with $p_{con}(\theta)$ (Eq.~\ref{eq:percprob}). The probability distribution of transport ability versus porosities for an ensemble of networks of resolution~$r$ then follows as: $P^{(r)}(\theta = x,ta =y) = P_\theta^{(r)}(x)\cdot p_{con}(x) \cdot P_\mathrm{ta}^{(r)}(y)$. 

Figure~\ref{fig:cloud_TA_pdf} shows the pdf-cloud (as continuous counterpart to Figure~\ref{fig:scatter_eff}a) for various input parameters of porosity statistics $\mu$ and~$\sigma$ and $a_r$ and $b_r$ depending on resolution $r$. 
The upper panel shows the isolated impact of the porosity distribution at the small resolution $r=2$. The range of potential porosity values reduces with decreasing standard deviation $\sigma$ (Figure~\ref{fig:cloud_TA_pdf}c) while transport ability remains a broad distribution for each porosity value. 

The combined effect of increasing scale on the $\theta-ta$-distribution through the reduction in value spreading is displayed in the botom row of Figure~\ref{fig:cloud_TA_pdf}. Here $ta$ statistics decrease similarly to those of $\sigma_r$ following the scaling behaviour identified in the numerical simulations: $b_r = \max\left(0.05,b_2/{(r/2)^{d/4}}\right)$.

\section{Discussion}

The upscaling scheme and CV/GCV graphs (Figure~\ref{fig:ens_stats_evolution}) allow determining a minimum sample size to be considered an REV for the material studied here. The procedure is transferable to other observations. In case the experimental equipment or computational resources are insufficient to handle an REV, the alternative is repeating the measurements sufficiently often at similar sub-REV domains. The  measurements will reveal a scatter with an unknown standard deviation. An appropriate sample size must be estimated during the process using standard statistical tools.

\subsection{REVs of FCC data}

Figure~\ref{fig:ens_stats_evolution} shows that 
The investigated FCC particles require a domain size of $512\,\mu$m to guarantee an REV for porosity at a level of $\alpha = 0.025$. However, FCC particles have an average diameter of $100\,\mu$m being below REV scale. The significant differences in the mean porosity of the $5$ particles reported in \citet{deWinter16} confirm that observations are below REV scale. It takes about $34$ individual particles of $100\,\mu$m diameter to cover the same surface area as an REV ($512\,\mu$m x $512\,\mu$m). Considering the spatial correlation of porosity and statistical soundness, the minimum number of particles will be much higher than these $34$ samples. 

Reaching REV level for transport ability $ta$ requires even larger surface areas when observing transport in a 2D setting. The domain scale of $512\,\mu$m would provide an acceptable level of deviation when investigating transport in 3D. When limited to particles of $100\,\mu$m diameter, $256$ particles are needed to cover the same volume. Even this absolute minimum number of particles is currently beyond the practical capabilities for effective parameters observations in lab studies.

\subsection{Implications for Continuum Equations and Experiments}

The observation that the GCV of $ta$ does not reach zero at large domain sizes has implications for the use of effective parameters in continuum equations.  A certain level of scatter remains in upscaled effective parameters. 
Accounting for this effect of pore space complexity in numerical and analytical solutions of continuum equations would require a representation of parameter variation through uncertainty bands and/or stochastic parameter representations which is hardly done.

Diffusion is probably the simplest form of porous media transport. Yet the scatter in $ta$ does not diminish fast nor completely and remains significant at relatively large length scales. This explains why fitted relations between an effective diffusion coefficient and porosity for porous materials vary among publications. Not even considering experimental errors, a small sample size/number can easily lead to fitting results which do not necessarily be representative for the material under investigation. 
Although other materials will reveal different statistical relations and scale-developments as presented here, we conclude that more focus should be given to a sufficient sample scale and sample number in experimental and computational studies aiming to determine effective parameters at Darcy scale. 
Given a limitation in data points, we strongly advocate the use of probability distributions for upscaling parameters and property relation rather then empirical deterministic fits.

\subsection{Perspectives}

Suppose we consider more complex processes than diffusion, such as solute transport by advection and dispersion, two-phase flow, evaporation, dissolution-precipitation, coupling of free flow over porous media flow, etc. Each process is influenced by the complexity of the pore space. Crucial open questions for future research remain: Do all processes need their own PDF for a complete description? And how do parameter PDF's develop during upscaling? What are appropriate REV scale given a certain process and porous material.

\section{Summary and Conclusions}

Applying the REV concept to transport parameters in porous media relies on the reduction of scatter with increasing observation scale to a negligible level. If scatter in observations persists, an effective mean is not a representative parameter for the process of interest. We investigated the scatter in observations of porosity and transport ability, being a parameter to represent the effect of porous medium structure on pore scale diffusion, with increasing scale from nanometer to micrometer. We made use of data obtained from FCC materials of \citet{deWinter16}.

The statistical analysis of the large collection of diffusive transport in sub-REV porous domains revealed the relationship between porosity and transport ability at multiple length scales. The assessment included the study of connected and disconnected volumes, where the latter inhibit diffusive flow at all.
We observe strong scatter of both, porosity and transport ability, as expected at sub-REV scale. Consequently, we relate porosity and transport ability through probability distribution functions (PDFs) instead of attempting to establish an explicit deterministic relation between both, which cannot exist due to the variability of the porous media structure. The PDFs provide a mathematical description of this variability. While porosity follows a normal distribution, transport ability is best characterized through a log-normal distribution. The statistical descriptions allowed performing numerical and theoretical upscaling.

The scatter in both quantities reduced with increasing scale, but with different convergence behaviour. We assess reaching the REV level for porosity by determining the coefficient of variation. The latter reduces linearly with increasing domain size due to a constant mean and the reduction in standard deviation. However, spatial correlation of porosity is observable in the data and reduces the convergence to an REV level. 

The scatter of the transport ability with increasing samples sized does not decrease linearly, but levels off at an asymptotic value $>0$. Using the Geometric Coefficient of Variation (GCV) the value is in the range of $0.05$. Thus, transport ability measurements will always show a scatter around the mean, regardless of the domain size. This can be easily explained by the natural variability of void space topology of porous media which will be present at all scale leading to a natural scatter in transport characteristics. However, even when ignoring the scatter and assuming the REV level is reached at GCV$ = 0.05$, the observation scale is in the range of $10000\,\mu$m$=1$\,cm. Thus, diffusion experiments at microscopic level in heterogeneous porous material must results in scattered observations not leading to representative results unless experiments are repeated for sufficient samples.

The influence of local heterogeneity to the transport behavior has consequences for the application of the REV concept which is key factor for (i) experimental work; (ii) continuum equations; (iii) simulations of large scale processes.

Our results lead us to the major conclusions: 
\begin{itemize}
    \item Although porosity observations converge to an effective value with increasing observation scale and/or number of samples; spatial correlation of samples lead to higher REV levels as typically assumed.
    \item The length scale of the REV for transport in porous media is easily underestimated when based on porosity measurements.    
    \item Diffusion parameters for microscopic samples will always show a scatter in measurements for heterogeneous porous material given the tortuous and non-unique pore space topology. 
    In the example of FCC particles, we found a minimum scatter of $5\%$. However, the level was only reached for large domain sizes leading to an REV level at cm scale.
    \item There is no one-to-one deterministic relationship between porosity and transport characterizing parameters, like the diffusion coefficient, in complex porous structures. Instead a probabilistic relation of these parameters to porosity is warranted to capture the effect of the complex pore space, particularly - but not exclusively - below REV level.
\end{itemize}

\section*{Declarations}

\paragraph{Funding:}
Matthijs de Winter is kindly supported by the Deutsche Forschungsgemeinschaft (DFG, German Research Foundation) – Project number 327154368 – SFB 1313.

\paragraph{Conflicts of interests:}
The authors declare that the research was conducted in the absence of any commercial or financial relationships that could be construed as a potential conflict of interest.

\paragraph{Availability of data and code:}
Manuscript data and codes are included as electronic supplementary material. 

\bibliographystyle{agu}
\bibliography{references}

\begin{thebibliography}{21}
\providecommand{\natexlab}[1]{#1}
\expandafter\ifx\csname urlstyle\endcsname\relax
  \providecommand{\doi}[1]{doi:\discretionary{}{}{}#1}\else
  \providecommand{\doi}{doi:\discretionary{}{}{}\begingroup
  \urlstyle{rm}\Url}\fi

\bibitem[{\textit{Bear}(1972)}]{Bear72}
Bear, J., \textit{Dynamics of Fluids in Porous Media}, Elsevier, New York,
  1972.

\bibitem[{\textit{Blunt et~al.}(2013)\textit{Blunt, Bijeljic, Dong, Gharbi,
  Iglauer, Mostaghimi, Paluszny, and Pentland}}]{Blunt13}
Blunt, M.~J., B.~Bijeljic, H.~Dong, O.~Gharbi, S.~Iglauer, P.~Mostaghimi,
  A.~Paluszny, and C.~Pentland, Pore-scale imaging and modelling, \textit{Adv.
  Water Resour.}, \textit{51}, 197--216, \doi{10.1016/j.advwatres.2012.03.003},
  2013.

\bibitem[{\textit{Bruckler et~al.}(1989)\textit{Bruckler, Ball, and
  Renault}}]{Bruckler89}
Bruckler, L., B.~C. Ball, and P.~Renault, Laboratory estimation of gas
  diffusion coefficient and effective porosity in soils, \textit{Soil Science},
  \textit{147}(1), 1--10, 1989.

\bibitem[{\textit{Cnudde and Boone}(2013)}]{Cnudde13}
Cnudde, V., and M.~N. Boone, High-resolution {X}-ray computed tomography in
  geosciences: {A} review of the current technology and applications,
  \textit{Earth-Sci Rev}, \textit{123}, 1--17,
  \doi{10.1016/j.earscirev.2013.04.003}, 2013.

\bibitem[{\textit{Dagan}(1989)}]{Dagan89}
Dagan, G., \textit{Flow and Transport on Porous Formations}, Springer, New
  York, 1989.

\bibitem[{\textit{de~Winter et~al.}(2016)\textit{de~Winter, Meirer, and
  Weckhuysen}}]{deWinter16}
de~Winter, D. A.~M., F.~Meirer, and B.~M. Weckhuysen, {FIB}-{SEM} tomography
  probes the mesoscale pore space of an individual catalytic cracking particle,
  \textit{ACS Catal.}, \textit{6}(5), 3158--3167,
  \doi{10.1021/acscatal.6b00302}, 2016.

\bibitem[{\textit{Ertl et~al.}(2008)\textit{Ertl, Knözinger, Schüth, and
  Weitkamp}}]{Ertl08}
Ertl, G., H.~Knözinger, F.~Schüth, and J.~Weitkamp (Eds.), \textit{Handbook
  of Heterogeneous Catalysis}, second ed., Wiley‐VCH, 2008.

\bibitem[{\textit{Ghanbarian et~al.}(2013)\textit{Ghanbarian, Hunt, Ewing, and
  Sahimi}}]{Ghanbarian13}
Ghanbarian, B., A.~G. Hunt, R.~P. Ewing, and M.~Sahimi, Tortuosity in porous
  media: {A} critical review, \textit{Soil Sci. Soc. Am. J.}, \textit{77}(5),
  1461--1477, \doi{10.2136/sssaj2012.0435}, 2013.

\bibitem[{\textit{Grunwaldt et~al.}(2013)\textit{Grunwaldt, Wagner, and
  Dunin‐Borkowski}}]{Grunwaldt13}
Grunwaldt, J.-D., J.~B. Wagner, and R.~E. Dunin‐Borkowski, Imaging catalysts
  at work: {A} hierarchical approach from the macro- to the meso- and
  nano-scale, \textit{ChemCatChem}, \textit{5}(1), 62--80,
  \doi{10.1002/cctc.201200356}, 2013.

\bibitem[{\textit{Hazen}(1893)}]{Hazen93}
Hazen, A., Some physical properties of sands and gravels: {With} special
  reference to their use in filtration, \textit{Tech. Rep. Twenty Fourth Annual
  Report}, State Board of Health of Massachusetts, 1893.

\bibitem[{\textit{Jayarathne et~al.}(2020)\textit{Jayarathne, Deepagoda,
  Clough, Nasvi, Thomas, Elberling, and Smits}}]{Jayarathne20}
Jayarathne, J. R. R.~N., T.~K. K.~C. Deepagoda, T.~J. Clough, M.~C.~M. Nasvi,
  S.~Thomas, B.~Elberling, and K.~Smits, Gas-{Diffusivity} based
  characterization of aggregated agricultural soils, \textit{Soil Sci. Soc. Am.
  J.}, \textit{84}(2), 387--398, \doi{10.1002/saj2.20033}, 2020.

\bibitem[{\textit{Jiang et~al.}(2013)\textit{Jiang, Dijke, Sorbie, and
  Couples}}]{Jiang13}
Jiang, Z., M.~I. J.~v. Dijke, K.~S. Sorbie, and G.~D. Couples, Representation
  of multiscale heterogeneity via multiscale pore networks, \textit{Water
  Resour. Res.}, \textit{49}(9), 5437--5449,
  \doi{https://doi.org/10.1002/wrcr.20304}, 2013.

\bibitem[{\textit{Karimpouli and Tahmasebi}(2016)}]{Karimpouli16}
Karimpouli, S., and P.~Tahmasebi, Conditional reconstruction: {An} alternative
  strategy in digital rock physics, \textit{Geophysics}, \textit{81}(4),
  D465--D477, \doi{10.1190/geo2015-0260.1}, 2016.

\bibitem[{\textit{Koltermann and Gorelick}(1996)}]{Koltermann96}
Koltermann, C.~E., and S.~M. Gorelick, Heterogeneity in sedimentary deposits: A
  review of structure-imitating, process-imitating, and descriptive approaches,
  \textit{Water Resour. Res.}, \textit{32}(9), 2617--2658,
  \doi{10.1029/96WR00025}, 1996.

\bibitem[{\textit{Kozeny}(1953)}]{Kozeny53}
Kozeny, J., \textit{Hydraulik: {Ihre} {Grundlagen} und praktische {Anwendung}},
  Springer, Wien, \doi{10.1007/978-3-7091-7592-7}, 1953.

\bibitem[{\textit{Mehmani et~al.}(2020)\textit{Mehmani, Kelly, and
  Torres-Verdín}}]{Mehmani20}
Mehmani, A., S.~Kelly, and C.~Torres-Verdín, Leveraging digital rock physics
  workflows in unconventional petrophysics: {A} review of opportunities,
  challenges, and benchmarking, \textit{Journal of Petroleum Science and
  Engineering}, \textit{190}, 107,083, \doi{10.1016/j.petrol.2020.107083},
  2020.

\bibitem[{\textit{S\'{a}nchez-Vila et~al.}(2006)\textit{S\'{a}nchez-Vila,
  Guadagnini, and Carrera}}]{SanchezV06}
S\'{a}nchez-Vila, X., A.~Guadagnini, and J.~Carrera, Representative hydraulic
  conductivities in saturated groundwater flow, \textit{Rev. Geophys.},
  \textit{44}, RG3002, \doi{10.1029/2005RG000169}, 2006.

\bibitem[{\textit{Schultze-Jena et~al.}(2020)\textit{Schultze-Jena, Boon,
  de~Winter, Bussmann, Janssen, and van~der Padt}}]{SchultzeJ20}
Schultze-Jena, A., M.~A. Boon, D.~A.~M. de~Winter, P.~J.~T. Bussmann, A.~E.~M.
  Janssen, and A.~van~der Padt, Predicting intraparticle diffusivity as
  function of stationary phase characteristics in preparative chromatography,
  \textit{J Chromatogr A}, \textit{1613}, 460,688,
  \doi{10.1016/j.chroma.2019.460688}, 2020.

\bibitem[{\textit{Shen and Chen}(2007)}]{Shen07}
Shen, L., and Z.~Chen, Critical review of the impact of tortuosity on
  diffusion, \textit{Chem Eng Sci}, \textit{62}(14), 3748--3755,
  \doi{10.1016/j.ces.2007.03.041}, 2007.

\bibitem[{\textit{van Brakel and Heertjes}(1974)}]{VanBrakel74}
van Brakel, J., and P.~Heertjes, Analysis of diffusion in macroporous media in
  terms of a porosity, a tortuosity and a constrictivity factor, \textit{Int J
  Heat Mass Transf}, \textit{17}(9), 1093--1103,
  \doi{10.1016/0017-9310(74)90190-2}, 1974.

\bibitem[{\textit{Zhang et~al.}(2000)\textit{Zhang, Zhang, Chen, and
  Soll}}]{Zhang00}
Zhang, D., R.~Zhang, S.~Chen, and W.~E. Soll, Pore scale study of flow in
  porous media: {Scale} dependency, {REV}, and statistical {REV},
  \textit{Geophys. Res. Lett.}, \textit{27}(8), 1195--1198,
  \doi{https://doi.org/10.1029/1999GL011101}, 2000.

\end{thebibliography}
\include{bibliography}

\end{document}


\maketitle

\section{Observation Data and Statistical Description}

\subsection{Statistical Analysis of Transport Ability Data}

We analysed the frequency of $ta$ values for the data set of $5128$ transport ability observations displayed in Figure~2 in the manuscript. They refer to diffusion simulation in cubic virtual volumes with a computer-generated pore space mapping the structure of the FFC particles at a length scale of $r=2 \mu m$ \citep{deWinter16}. Note that \cite{deWinter16} analysed transport resistance $tr$ which is related to transport ability through $tr = 1/ta$. Their figure 6b corresponds to Figure~3 in the manuscript. \cite{deWinter16} further assumed a simple truncated normal distribution function for the relation between $ta$ and porosity $\theta$. We refined the analysis with through statistically sound normality and log-normality testing.

We determine the characteristics of the probability distribution $P_{ta}(\theta)$ through statistical frequency analysis. We grouped data into porosity bins of width $0.02$. For statistical soundness, bins with less then $20$ values are not further considered. For each bin, the frequency distribution of transport ability values $F_{ta}(\theta)$ is interpreted as the probability $P_{ta}(\theta)$ of a $ta$-value as a particular porosity value $\theta$. 

For each porosity value, we performed statistical hypothesis testing on normality and log-normality using the Shapiro-Wilk Test, D’Agostino’s Test and the Anderson-Darling Test at significance level of $5\%$. The tests indicated that frequency distributions $F_{ta}(\theta)$ are log-normal distributed for small values of porosity $\theta_\mathrm{con}<0.2$ and rather normal distributed for connected porosities around the mean porosity $\theta \approx \mu \approx 0.3$.  For large values of porosity ($\theta>0.5$), the transport ability values fall into a small range of high transport ability values not giving positive outcomes of the normality tests for neither scales. Results are exemplified for two porosity values in Figure~3 in the manuscript.

We attribute the tendency of $F_{ta}(\theta)$ towards a log-normal distribution for small porosities to the increased impact of pore space topology. The scarce pore space is either scattered with several bottlenecks or forms one or few flow channels. In the first domain type, flow is hampered and $ta$ will be low, while in the latter we observe quasi plug flow with a relatively high $ta$ values. The significant tail of low $ta$ values refers to domains which are hardly connected. With increasing porosity more connected pathways exist through the network. The structural differences between individual volumes reduce, and $ta$ tends to be rather normal distributed. 

We calculated expectation values $a_{\log ta}$ and standard deviations $b_{\log ta}$ for each log-transformed frequency distribution $F_{ta}(\theta_\mathrm{con})$. The log-transformation allow a proper comparison of statistics given the logarithmic nature of $ta$ values. The log-normal distributions also represents the frequency for higher porosities, as it converges to the normal-distribution for decreasing standard deviations. Results are displayed in Figure~4 in the manuscript.
The log-$ta$ mean $a(\theta)$ shows a linear relationship for porosities smaller then $0.6$ and then a flattening towards $0$ since $ta(\theta=1) = 1$. $\log-ta$ standard deviation is high for small porosities and decreases with increasing value. For porosities beyond $0.6$, the scatter in $ta$ is negligible. 

Transport ability $ta$ as stochastic function of porosity follows from the statistical descriptions of connectivity (Eq.~5, Manuscript) and connected transport ability (Eq.~4, Manuscript) with
\begin{equation}\label{eq:prob_ta}
P_\mathrm{ta}(ta = y,\theta) = 
\begin{cases}
    0  & \text{ with } 1-p_\mathrm{con}(\theta) \\
    y>0 & \text{ with } p_\mathrm{con}(\theta)\cdot P_\mathrm{ta}^\textup{con}(y,\theta)
\end{cases}
\end{equation}

The expectation value of $ta$ follow as $\mathrm{E}[ta](\theta) = \mathrm{e}^{a(\theta) + b^2(\theta)/2}\cdot p_\mathrm{con}(\theta)$ following from the first moment of the log-normal distribution and the fact that disconnected elements  contribute $ta=0$ by the amount of $1-p_\mathrm{con}$. The variance and $\mathrm{VAR}[ta](\theta) = p_\mathrm{con}(\theta)\cdot \mathrm{e}^{2a(\theta) + b^2(\theta)}\left(\mathrm{e}^{b^2(\theta)} -  p_\mathrm{con}(\theta)\right)$ which follows from the second moment of the log-normal distribution in combination with the probability of connected elements.
Theoretical results of expectation value refers to ergodic conditions, i.e. when the domain size reaches REV level for all involved processes. 

\section{Upscaling}

\subsection{Numerical Upscaling Procedure}

\begin{figure}[ht]
\centering
    \includegraphics[width=0.5\textwidth]{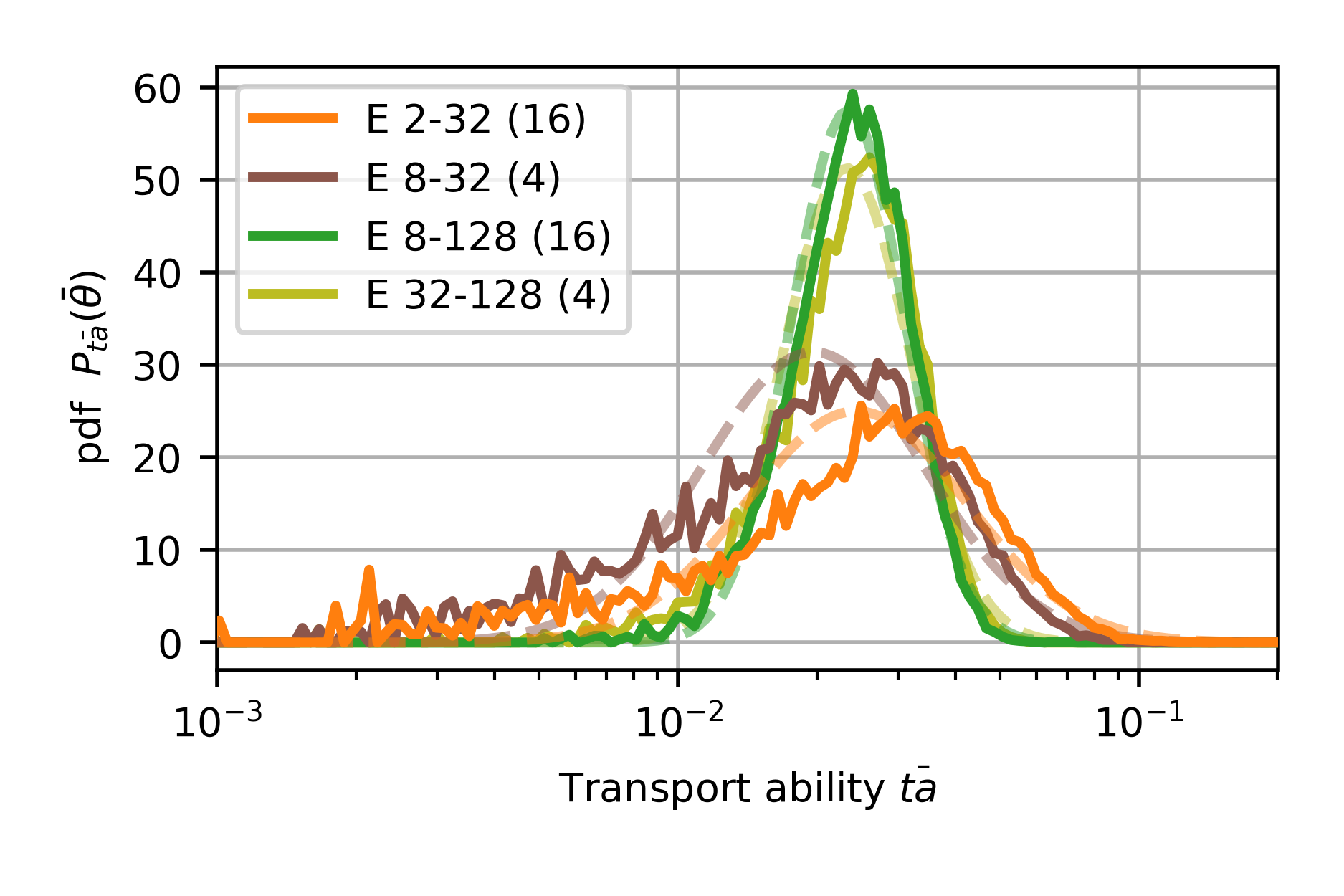}
    \caption{Impact of upscaling step size: Upscaled ensemble transport ability pdf for final resolution of $32\,\mu$m and $128\,\mu$m archived by upscaling at domain increase step sizes $n=4$ and $n=16$.}
    \label{fig:ta_pdf_resolution}
\end{figure}

Networks consist of $N = n^d$ nodes, with $d$ being the dimension and $n$. We decided for $n = 16$
sub-samples per direction based on network theory and preliminary tests showing that boundary effects are
negligible for $n > 10$. Figure~\ref{fig:ta_pdf_resolution} shows the results of preliminary tests on sub-samples size. We generated $4$ ensembles:
\begin{itemize}
    \item E{2-32}: upscaling from $2\mu$m to $32\mu$m resolution with $n=16$ subsamples per direction;
    \item E{8-32}: upscaling from $8\mu$m to $32\mu$m resolution with $n=4$ subsamples per direction;
    \item E{8-128}: upscaling from $8\mu$m to $128\mu$m resolution with $n=16$ subsamples per direction;
    \item E{32-128}: upscaling from $32\mu$m to $128\mu$m resolution with $n=4$ subsamples per direction.
\end{itemize}

The similar $ta$ distributions for resulting resolutions  confirm that the upscaling step size is negligible. This also supports the use of upscaling from $2\,\mu$m to a resolution of $r = 8\,\mu$m in a stepsize of $n = 4$ elements per dimension.


\subsection{Numerical Upscaling Results}
\begin{figure}[ht]
\centering
    \includegraphics[width=0.475\textwidth]{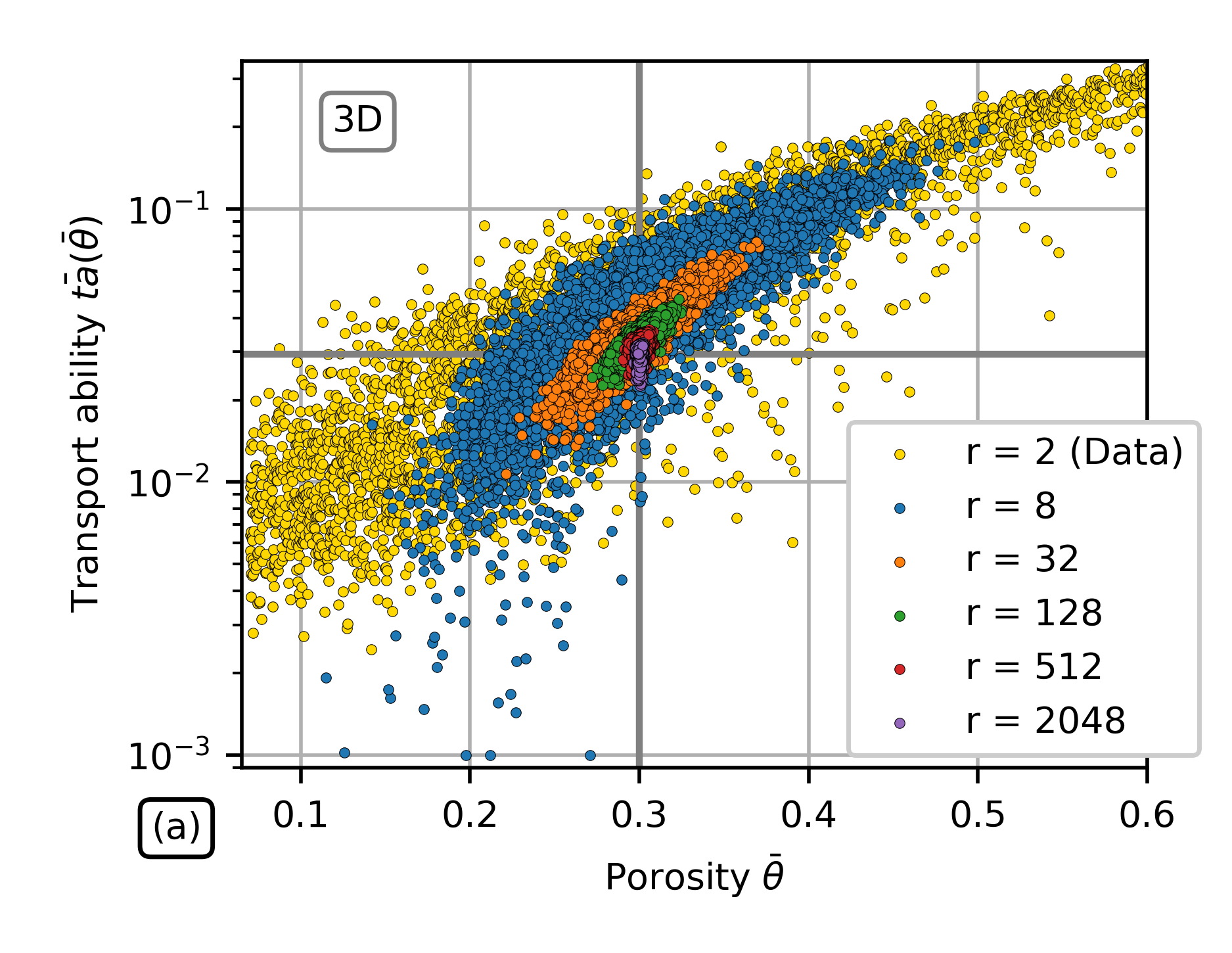}
        \includegraphics[width=0.95\textwidth]{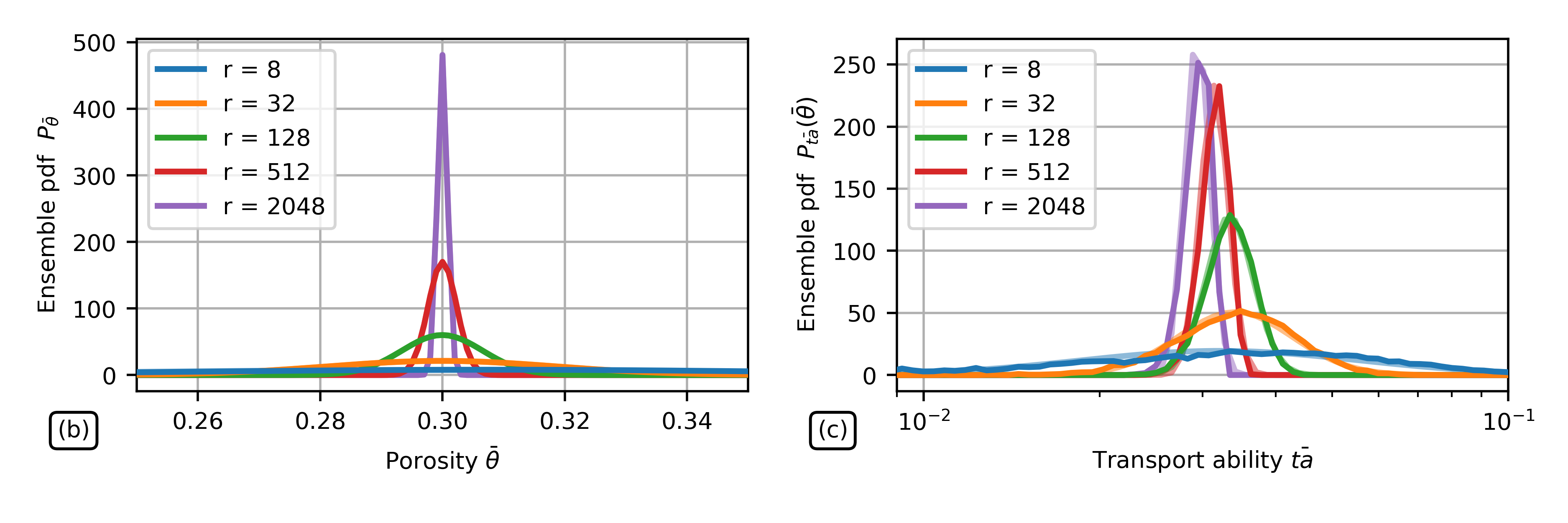}
        \caption{Results of numerical upscaling for simulated ensembles at different resolutions $r$ [$\mu$m] in 3D: (a) Scatter of ensemble transport ability $\bar{ta}$ vs porosity $\bar\theta$; (b) and (c) marginal distributions (i.e. normalized histograms) for porosity and transport ability. The lighter lines in (b) show the associated log-normal distributions based on mean and standard deviations of the log-$ta$ ensemble values.}
    \label{fig:scatter_eff}
\end{figure}

Figure~\ref{fig:scatter_eff} displays the numerical upscaling results for domain extension in 3D as addition to the results shown for 2D extension presented in Figure~6 in the manuscript.

\bibliographystyle{agu}
\bibliography{references}
\include{bibliography}